\documentclass[iop]{emulateapj}
\usepackage{natbib}
\usepackage{graphicx}
\usepackage{hyperref}
\usepackage{breakurl}
\usepackage{apjfonts}
\usepackage{longtable}
\usepackage{gensymb}
\usepackage{amsmath}
\usepackage{listings}
\usepackage{graphicx}
\usepackage{epstopdf}
\usepackage{grffile}

\begin{document}

\title{ON THE EVOLUTION OF MAGNETIC WHITE DWARFS}

\author{P.-E.~Tremblay$^{1,2}$, G. Fontaine$^{3}$, B.~Freytag$^{4}$,
   O. Steiner$^{5,6}$, H.-G. Ludwig$^{7}$, M.~Steffen$^{8}$, S. Wedemeyer$^{9}$, and P. Brassard$^{3}$}

\affil{$^{1}$Space Telescope Science Institute, 3700 San Martin Drive,
  Baltimore, MD, 21218, USA} \affil{$^{2}$Hubble Fellow}
\email{tremblay@stsci.edu} \affil{$^{3}$D\'epartement de Physique,
  Universit\'e de Montr\'eal, C. P. 6128, Succursale Centre-Ville,
  Montr\'eal, QC H3C 3J7, Canada} \affil{$^{4}$Department of Physics
  and Astronomy at Uppsala University, Regementsv\"agen 1, Box 516,
  SE-75120 Uppsala, Sweden} \affil{$^{5}$Kiepenheuer-Institut f\"ur Sonnenphysik,
  Sch\"oneckstr. 6, D-79104 Freiburg, Germany} \affil{$^{6}$Istituto Ricerche Solari Locarno, Via Patocchi 57, 6605 Locarno-Monti, Switzerland} \affil{$^{7}$Zentrum f\"ur Astronomie der
  Universit\"at Heidelberg, Landessternwarte, K\"onigstuhl 12, D-69117
  Heidelberg, Germany} \affil{$^{8}$Leibniz-Institut f\"ur Astrophysik
  Potsdam, An der Sternwarte 16, D-14482 Potsdam, Germany}
 \affil{$^{9}$Institute
  of Theoretical Astrophysics, University of Oslo, PO Box 1029
  Blindern, N-0315 Oslo, Norway}

\begin{abstract}

We present the first radiation magnetohydrodynamics simulations of the atmosphere
of white dwarf stars. We demonstrate that convective energy transfer
is seriously impeded by magnetic fields when the plasma-$\beta$ parameter, the thermal to magnetic 
pressure ratio, becomes smaller than unity. The critical field strength
that inhibits convection in the photosphere of white dwarfs is in the 
range $B=1$-50~kG, which is much smaller than the typical 1-1000~MG field strengths observed in magnetic white dwarfs,
implying that these objects have radiative atmospheres. We have then
employed evolutionary models to study the cooling process of high-field magnetic white dwarfs, where
convection is entirely suppressed during the full evolution ($B \gtrsim 10$~MG). We find that the inhibition of convection
has no effect on cooling rates until the effective temperature ($T_{\rm eff}$) reaches a value of around 5500~K. 
In this regime, the standard convective sequences start to deviate from the ones without convection
owing to the convective coupling between the outer layers and the degenerate reservoir of thermal
energy. Since no magnetic
white dwarfs are currently known at the low temperatures where
this coupling significantly changes the evolution, effects of magnetism on cooling rates are 
not expected to be observed.
This result contrasts with a recent suggestion that magnetic 
white dwarfs with $T_{\rm eff} \lesssim 10,000$~K cool significantly slower
than non-magnetic degenerates.

\end{abstract}


\section{INTRODUCTION}

Magnetic white dwarfs are stellar remnants featuring global magnetic
structures with field strengths from 1~kG to 1000~MG. They
account for a significant part of the white dwarf population with an
estimated fraction of around 10\% in volume complete samples
\citep{liebert03,schmidt03,kawka07}. Most of these objects are
high-field magnetic white dwarfs (HFMWD), with field strengths $B >
1$~MG, and a distribution of magnetic field strengths that appears to peak
around $\sim 20$~MG \citep{schmidt03,kulebi09}. HFMWDs show obvious
Zeeman line splitting in spectroscopic observations. It is currently
difficult to understand these data owing to the lack
of an appropriate theory of Stark broadening in the presence of a
background magnetic field in an arbitrary direction \citep{main98}.
In particular, the standard spectroscopic technique employed to derive atmospheric
parameters from the Balmer lines \citep{bergeron92} can not be used
to constrain the mass and cooling age of HFMWDs
\citep{kulebi09}. However, there is a growing sample of HFMWDs in common 
proper motion pairs or with known trigonometric parallaxes, 
allowing to derive masses. This sample shows a mean mass of $\sim 0.80$~$M_{\odot}$ \citep{ferrario15,briggs15}, 
which is significantly higher than the mean mass of non-magnetic 
white dwarfs \citep[$\sim 0.60$~$M_{\odot}$, see,
  e.g.,][]{kleinman13}.

Numerous recent studies have provided scenarios for the origin of
HFMWDs, accounting for their mass, velocity, and magnetic field strength distributions. The
lack of a significant trend in the number of HFMWDs as a function of
atmospheric composition and cooling age \citep{kulebi09}, as well the presence of field
strengths too large to be produced by a convective dynamo \citep{dufour08}, suggest
that magnetic fields are remnants of the white dwarf progenitors. 
Current scenarios tend to be grouped in three
categories, suggesting that HFMWDs are: remnants of intermediate mass
stars with conserved fossil fields \citep{angel81,wickra05}; the
outcome of mergers of either two white dwarfs or a white dwarf and the
core of a giant star \citep{garcia12,kulebi13a,wickra14,briggs15}; or
products of the amplification of a seed field by a convective dynamo
in the core of the evolved progenitors
\citep{ruderman73,kissin15}. The origin of magnetic white dwarfs
remains elusive since current observations do not allow to clearly 
differentiate between these evolution channels. Magnetic white dwarfs in clusters and common
proper motion pairs \citep{kulebi10,dobbie12,kulebi13b,dobbie13}
are consistent with single star evolution in some but not all
cases. It is also difficult to explain the white dwarfs with the strongest magnetic fields, as
well as the absence of HFMWDs with late-type star companions, without
invoking the merger scenario \citep{wickra14, kissin15}.

On the other hand, weaker magnetic fields ($B \lesssim 1$ MG)
are also found in white dwarfs, although they are difficult to detect
systematically owing to the lack of obvious Zeeman splitting in high-resolution 
spectra for $B \lesssim$ 20 kG \citep{jordan07}. A few small
spectropolarimetric surveys, sensitive to field strengths as small as $\sim$1~kG, 
have put the fraction of kG-range white dwarfs at 3-30\%, with a roughly constant 1-10\% incidence per decade
of magnetic field strength \citep{jordan07,landstreet12,kawka12}. The uncertain
ratio reflects the small number statistics in the current
samples. These magnetic white dwarfs appear to have average masses and
are thought to origin from single stellar evolution \citep{jordan07},
with the magnetic fields possibly generated through a dynamo process
in the white dwarf progenitor \citep{wickra14}. Furthermore, large
observed values for $\langle B_{\rm z} \rangle/\langle
\vert B \vert \rangle$ are a strong indicator that the fields have 
a global organized structure, unlike the complex magnetic fields 
at the surface of Sun-like stars \citep{landstreet12}.

Gaia will provide precise parallaxes for more than 100,000 white
dwarfs, including all known magnetic white dwarfs
\citep{torres05,carrasco14}, and spectroscopic follow-ups
will identify even more magnetic objects. Gaia will establish
the first homogeneous mass distribution and cooling sequence 
of magnetic remnants. Given the ubiquitous presence of magnetic white dwarfs 
in the high-mass regime, it is critical to understand these objects to
recover the Galactic star formation history and initial mass function in the 
$\sim 3$-8~$M_{\odot}$ range \citep{tremblay14}. Magnetic remnants can also be used to constrain 
stellar evolution at intermediate masses \citep{kulebi13b} and study possible populations of 
mergers \citep{merger1,merger2}. It is therefore essential, at this stage,
to build precise model atmospheres and evolution sequences
for these peculiar degenerate stars. It has been suggested for a long time
that convection is completely inhibited in HFMWDs
\citep{wickra86,nature}, although this has not yet been verified with
realistic simulations. Furthermore, \citet{kepler13}
suggest that small undetected magnetic fields could impact
the mass distribution of cool convective white dwarfs. 

\citet{nature} have recently proposed that the inhibition of
convection in magnetic white dwarfs has a large impact on cooling
rates, by increasing the cooling times by a factor of two to three. 
However, they arrived at this conclusion with a simple analytical
argument and it needs to be verified with state-of-the-art 
evolution models. In this work, we perform the first radiation magnetohydrodynamics (RMHD) simulations
of the atmosphere of magnetic pure-hydrogen (DA) white dwarfs (Section 2.1). We then consider the
results of these simulations for the computation of new cooling sequences
for magnetic white dwarfs using an established evolution code (Section 2.2).
We discuss the implications of these results in Section 3 and conclude in Section 4.
 
\section{WHITE DWARF MODELS}

\subsection{Magnetohydrodynamics Simulations}

We have computed RMHD simulations with the CO$^{5}$BOLD code \citep{freytag12} 
for pure-hydrogen DA white dwarfs. We rely on a representative set of atmospheric parameters,
$T_{\rm eff} \sim 10,000$~K and a surface gravity of $\log g = 8.0$, and our simulations are detailed 
in Table~1. The setup of the 
simulations is very similar to that of the non-magnetic models presented
in \citet{tremblay13a,tremblay13}. In the temperature regime considered
here, the convection zone is significantly deeper than the atmospheric layers, 
and we use a bottom boundary (at Rosseland optical depth $\tau_{\rm R} \sim 10^{3}$) open to convective flows and radiation, where a 
zero net mass flux is ensured. We fix the entropy of the ascending material
at 2.0819 erg g$^{-1}$ K$^{-1}$ for all simulations, corresponding to the 
value in the non-magnetic simulation at $T_{\rm eff} = 10,025$ K and $\log g = 8.0$
from \citet{tremblay13}. The lateral boundaries are periodic, and the top boundary is open 
to material flows and radiation. We rely on the same opacities, equation of state, 
and grid resolution ($150 \times 150 \times 150$) as the previous non-magnetic simulations. 
The frequency dependent, i.e. non-gray, radiative transfer
is solved along long characteristics employing Feautrier's method. 
Opacities are grouped into bins (for details on the opacity binning 
approach see, e.g., \citealt{nordlund82,ludwig94,voegler04}) 
 using the 8-bin scheme of \citet{tremblay13}.

The main difference compared to earlier computations is that we have imposed, at the start of the simulations, a vertically 
oriented magnetic field (towards the exterior of
the star) with an amplitude of 0.5~kG and 5~kG for our two magnetic simulations, respectively. The magnetic
boundary conditions are imposed independently to the hydrodynamics conditions. We
require that magnetic field lines remain vertical at both the top and bottom
layers, while lateral boundaries are periodic. We further require that the magnetic flux 
is constant at the bottom, mimicking the effect of a global fossil field anchored in the deep
degenerate core. We note that our RMHD simulations do not assume hydrostatic equilibrium and 
automatically take into account the turbulent pressure, magnetic tension forces, and magnetic pressure.

\begin{deluxetable*}{lccccccccc}[!t]
 \tablecolumns{8}
 \tablewidth{0pt}
 \tablecaption{RMHD Models of White Dwarfs}
 \tablehead{
 \colhead{$T_{\rm eff}$} &
 \colhead{$\log g$} &
 \colhead{$x$} &
 \colhead{$z$} &
 \colhead{$z (\tau_{\rm R} = 1) - z_{\rm bot}$} &
 \colhead{$B_{\rm z}$} &
 \colhead{$B_{\rm z, rms} (\tau_{\rm R} = 1)$} &
 \colhead{$\delta I_{\rm rms}/\langle I \rangle$} &
 \colhead{Mach$(\tau_{\rm R} = 1)$} & \\
 \colhead{(K)} &
 \colhead{(cm s$^{-2}$)} &
 \colhead{(km)} &
 \colhead{(km)} &
 \colhead{(km)} &
 \colhead{(kG)} &
 \colhead{(kG)} &
 \colhead{(\%)} &
 \colhead{} &
 }
 \startdata
10024 & 8.0 & 2.11 & 0.83 & 0.38 & 0   & 0    & 14.69 & 0.45 \\
10037 & 8.0 & 2.11 & 0.83 & 0.38 & 0.5 & 1.28 & 14.13 & 0.38 \\
9147  & 8.0 & 2.11 & 0.63 & 0.30 & 5.0 & 5.38 & 21.86 & 0.25 \\
 \enddata
\tablecomments{All quantities were averaged over 250 snapshots and
over constant geometrical depth when appropriate. $T_{\rm eff}$ is derived from the temporal and spatial average of the emergent flux. $B_{\rm z}$ is the horizontally averaged 
magnetic field, which is constant at all times and all depths from the requirement of magnetic flux conservation. 
$B_{\rm z, rms}$ is the rms vertical magnetic field at the geometrical depth that
 corresponds to $\langle \tau_{\rm R} \rangle_{\rm x,y} = 1$. $\delta I_{\rm rms}/\langle I \rangle$ is the relative intensity 
contrast \citep[see Eq. (73) of][]{freytag12}.}
\label{tb:3D}
\end{deluxetable*}

The MHD module of CO$^{5}$BOLD \citep[see Section 3.7 of][]{freytag12} provides several numerical methods for solving the MHD equations, which are quite different from the ones
employed for pure hydrodynamics. In particular, the method used here relies on the HLL solver \citep{HLL},
which is more stable but with increased dissipation compared to the Roe solver used for the published grid of DA
white dwarfs. The MHD module also enforces the divergence-free 
condition $ \nabla \cdot B = 0$ based on a constrained transport scheme \citep[see, e.g.,][]{toth00}. 
In order to study the effect of magnetic fields on the
atmospheric stratification, we have computed a third model with the same MHD solver 
but no magnetic field. We computed all simulations for five seconds in stellar time, 
which is several times the convective turnover timescale.

\begin{figure*}[t]
\begin{center}
\includegraphics[width=0.33\linewidth,bb=100 400 450 730]{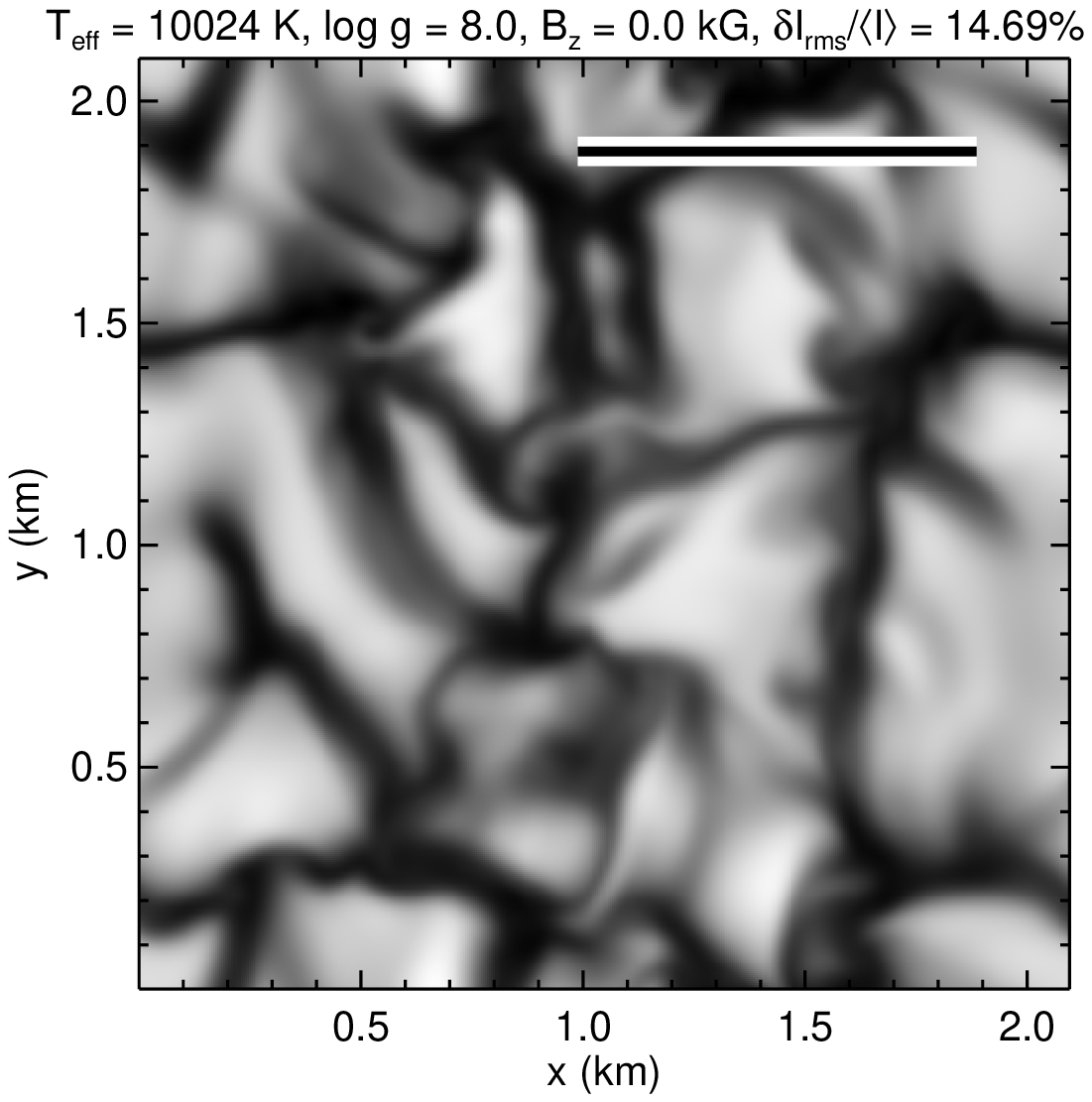}
\includegraphics[width=0.33\linewidth,bb=100 400 450 730]{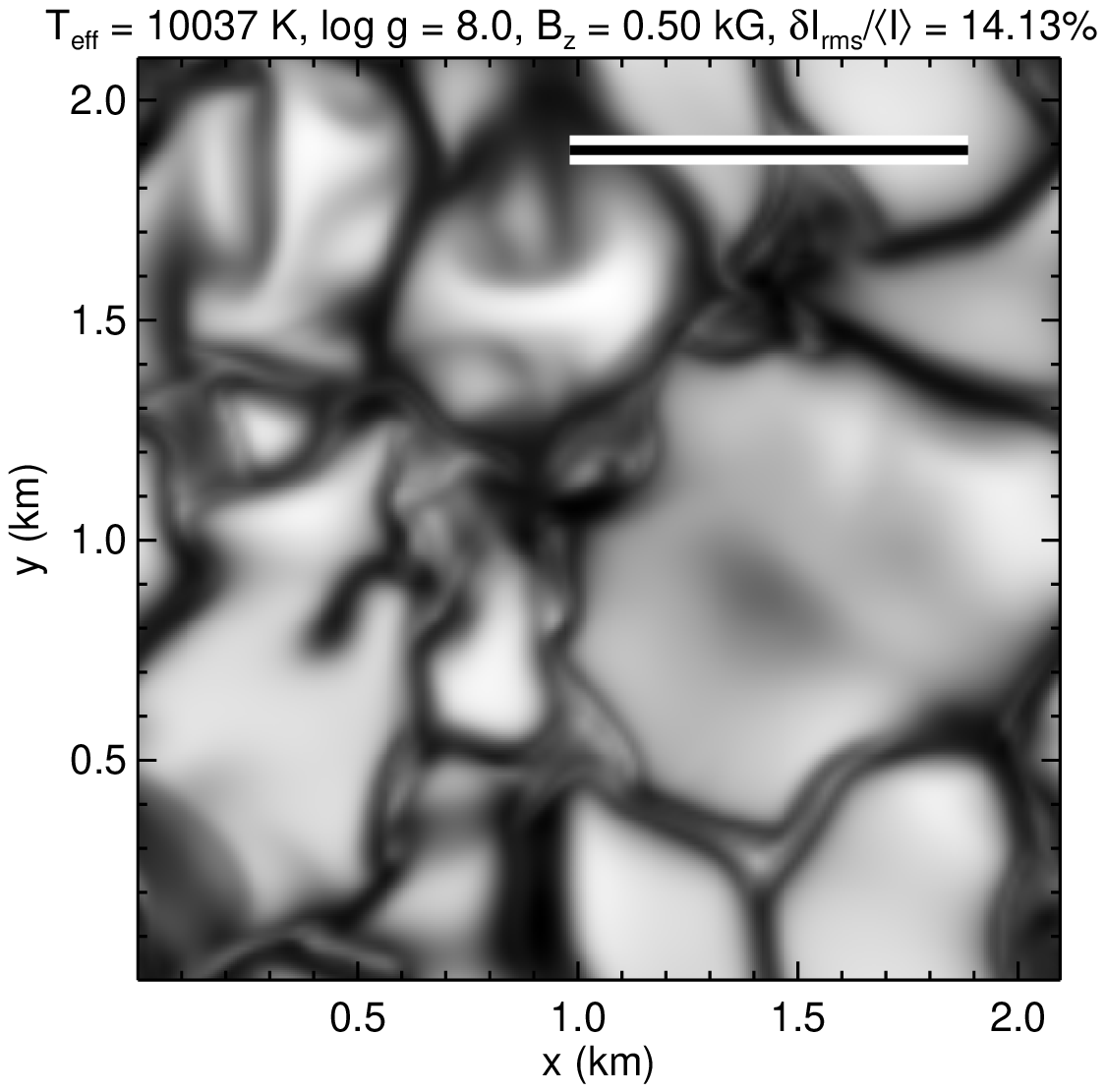}
\includegraphics[width=0.33\linewidth,bb=100 400 450 730]{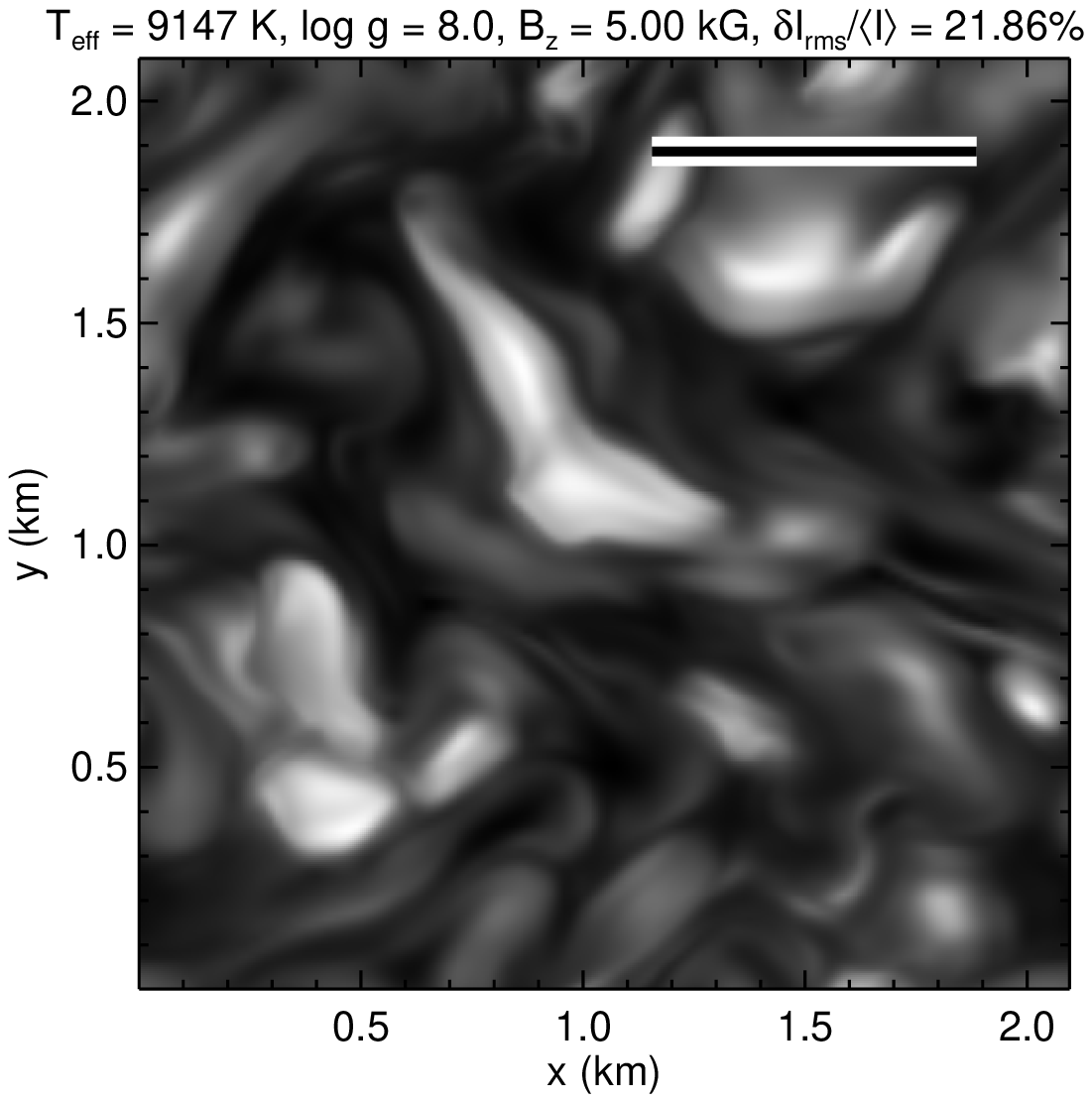}
\end{center}
\caption{Bolometric intensity emerging from
the $xy$ plane at the top of computational domain for CO$^{5}$BOLD 3D simulations computed with the MHD
	solver. All simulations have a constant surface gravity of $\log g = 8.0$, a pure-hydrogen composition, 
	and rely on the same entropy value 
	for the inflowing material through the open bottom boundary. The simulations shown in the middle
  and right panels have at the bottom an imposed average vertical magnetic field of
  0.5~kG and 5~kG, respectively. The rms intensity contrast with respect to the mean
  intensity and $T_{\rm eff}$ values are also shown above the panels. The length of
  the bar in the top right is 10 times the pressure scale height at
  $\tau_{\rm R} = 2/3$.
 \label{fg:f_int}}
\end{figure*}

Figure~\ref{fg:f_int} presents snapshots of the emergent intensity for
our three relaxed simulations. From an average over 125 snapshots, we also
display at the top of the panels the $T_{\rm eff}$ values (derived from the emergent flux) and the
relative intensity contrast. We observe that magnetic
fields have a significant impact on the emergent intensity. For
$B_{\rm z}=0.5$~kG, diverging upflows concentrate
magnetic flux in downflows, much like what is observed 
in the so-called quiet regions on the Sun \citep{nordlund09}, which are characterized by a rather weak average magnetic flux. Small magnetic flux concentrations form and appear as bright intergranular points since they act as radiative leak due to their reduced mass density. Table~1 demonstrates that the root-mean-square vertical magnetic field in the photosphere is significantly larger than the average magnetic field owing to these flux concentrations. For a field strength of $B_{\rm z}=5$~kG, convection is already largely inhibited, and occurs as narrow and bright plumes
very similarly to Sun spots where $B \gtrsim 2.5$~kG \citep{weiss96,sunspot1}.
This is not a surprising result since the thermal pressure in 
the photosphere of the simulated white dwarfs is only slightly larger than that
in the Sun, and a similar magnetic pressure is necessary to inhibit
convective flows. Studies of the impact of magnetic fields on surface convection 
in the Sun and Sun-like stars by numerous RMHD simulations \citep{rempel09,cheung10,freytag12,beeck13,steiner14}
can also be used to learn about the same process in white dwarfs,
even though the origin and large scale structure of magnetic fields
are very different.

\begin{figure}[t]
\begin{center}
\includegraphics[width=0.55\linewidth,bb=150 255 480 650]{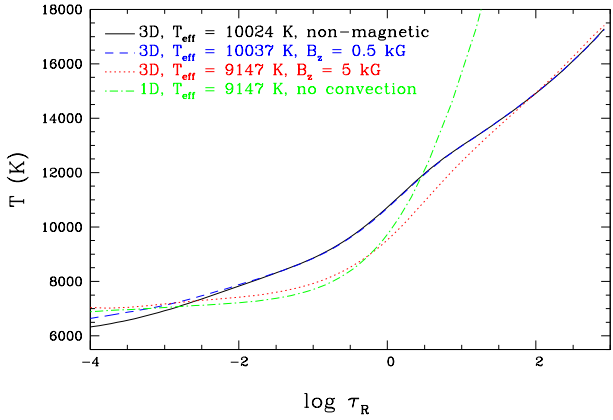}
\end{center}
\caption{Temperature structure as function of
optical depth (logarithmic scale) for the non-magnetic (black), 0.5~kG (blue), and 5~kG (red)
3D RMHD simulations. The temperature was determined from an average
of $\langle T^{4} \rangle$ over surfaces of constant optical depth. 
We also show a purely radiative 1D model atmosphere (green), where convection was artificially suppressed, 
at the same $T_{\rm eff}$ value as the 5~kG model.
 \label{fg:f_struc}}
\end{figure}

Figure~\ref{fg:f_struc} presents the temperature profiles of our
simulations, drawn from the average of $\langle T^{4}\rangle$ over surfaces of constant $\tau_{\rm R}$ for 12 snapshots. For
the 0.5~kG simulation, we observe that the magnetic field only
has an impact on the upper photosphere ($\tau_{\rm R} < 10^{-2}$),
where the temperature gradient is shallower. The importance
of the feedback effect of magnetic fields on the stellar structure can be
estimated from the plasma-$\beta$ parameter

\begin{equation}
\beta = \frac{8 \pi P}{B^2} ~,
\end{equation}

{\noindent}the thermal to magnetic pressure ratio, where $P$ is the
thermal pressure, and $B$ the average magnetic field strength. Since the
thermal pressure is rapidly decreasing with height while the magnetic
pressure is roughly constant, magnetic feedback effects increase
with height. There are two main reasons for the shallower temperature
gradient in the uppermost stable layers of magnetic white dwarfs. First of all, magnetic field lines 
restrain convective flows, hence the convective overshoot that usually
cools the upper layers is weaker \citep{tremblay13a}. This is a purely dynamical effect that will not be
observed in a 1D magnetic model with local convection. Furthermore,
the radiative heating, magnetic dissipation, and magnetic pressure all contribute
to increase the thermal pressure scale height compared to the non-magnetic case, 
which implies a shallower temperature gradient as a function of geometrical depth.
In general, the consequence is also a shallower temperature gradient as a function
of $\tau_{\rm R}$.

\begin{figure}[t]
\begin{center}
\includegraphics[width=0.55\linewidth,bb=150 255 480 650]{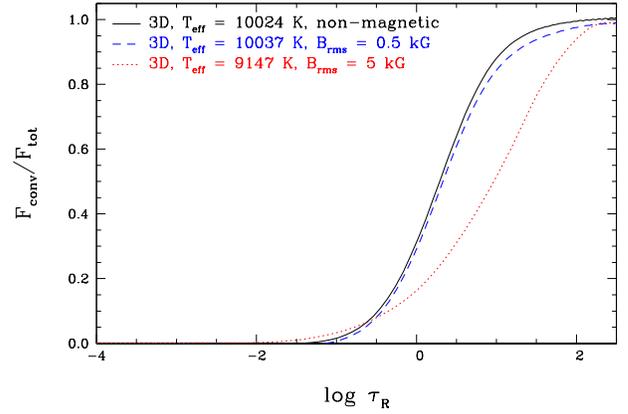}
\end{center}
\caption{Ratio of the convective to total energy flux as a function of 
optical depth (logarithmic scale) for the non-magnetic (black), 0.5~kG (blue), and 5~kG (red)
3D RMHD simulations. The $\langle$3D$\rangle$ convective flux is the sum of
the enthalpy and kinetic energy fluxes \citep[see Eq. 5 of][]{tremblay15}
averaged over constant geometrical depth.
 \label{fg:f_conv}}
\end{figure}

For a field strength of 5~kG, the overall atmospheric stratification is
significantly impacted by the presence of a magnetic field. Convective
energy transfer is impeded in the photosphere and Figure~\ref{fg:f_conv}
demonstrates that the convective flux at $\tau_{\rm R}$ = 1 is reduced by a factor of two compared
to the non-magnetic model. The smaller convective energy transfer implies that the stratification in the convectively unstable regions must adjust to a steeper temperature 
gradient to transport the same amount of total flux. The temperature gradient in the 
line-forming regions becomes very close to the radiative gradient, as demonstrated
in Figure~\ref{fg:f_struc} from the comparison with a 1D structure where convection 
was artificially suppressed. 
On the other hand, in the deeper layers where the 
thermal energy is larger, convection is still significant for this field strength.
Nevertheless, the steeper temperature gradient in the upper convectively unstable layers ($\tau_{\rm R} \gtrsim 0.1$), 
caused by the inhibition of convection, decreases the $T_{\rm eff}$ value by 880~K for 
the same conditions at the bottom. Full evolutionary calculations are necessary to link the magnetic atmospheres to 
the stellar interior, and this result does not imply that magnetic
white dwarfs have smaller luminosities for the same core temperature (see Section 2.2). 
For our models at $T_{\rm eff} \sim$ 10,000 K and $\log g$ = 8.0, $\beta$ = 1
for $B \sim$ 5.7 kG at the photosphere ($\tau_{\rm R}$ = 1). This critical field strength is very close to 
the observed transition between a convective and an almost fully
radiative temperature gradient in the RMHD simulations. Our results support the suggestion that 
when the plasma-$\beta$ is smaller than unity, i.e. when the magnetic pressure dominates over the thermal energy, the white dwarf atmospheric stratification adjusts to a radiative gradient since convective energy transfer is significantly
hampered.

In those cases where the plasma-$\beta$ parameter is smaller than unity, 
the atmosphere is not expected to become static or homogeneous since
the stratification is still convectively unstable, albeit unable to 
create energetically efficient convective flows. In particular, the relative 
intensity contrast for the $B$ = 5~kG simulation
is still 21.9\%, an even larger value than for the non-magnetic simulation. 
While convection is restricted to narrow and inefficient plumes, the temperature 
contrast and velocities in those structures are still large. It is 
currently unclear how these fluctuations would decrease as the magnetic
field strength is further increased. It is a serious technical challenge
to compute RMHD simulations with larger field strengths since the time steps
are dictated by the Alfv\'en speed $B/\sqrt{4\pi\rho}$, where $\rho$
is the density. For instance, the simulation at 5~kG is already of the order
of 10 times slower than the non-magnetic simulation. Finally, the 
magnetic field tends to form localized flux concentrations in the 
intergranular lanes, and the spatial resolution of our RMHD simulations
likely needs to be improved in order to properly characterize the
intensity contrast and small-scale fluctuations.

\begin{figure}[t]
\begin{center}
\includegraphics[width=0.55\linewidth,bb=150 255 480 645]{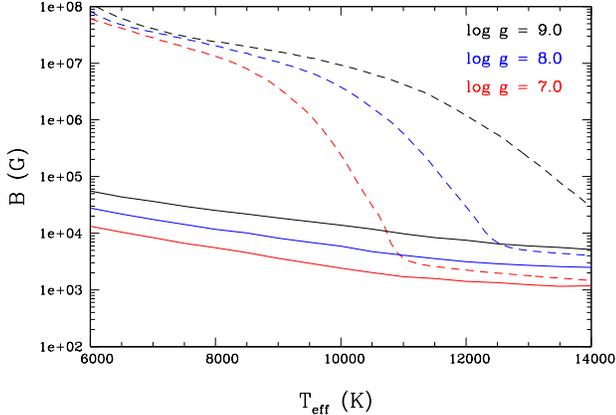}
\end{center}
\caption{Magnetic field strength that corresponds to plasma-$\beta = 1$
in the photosphere ($\tau_{\rm R} = 1$, solid lines) and base
of the convection zone (dashed lines) as a function of $T_{\rm eff}$.
Sequences are color-coded for $\log g$ = 7.0 (red), 8.0 (blue), and 9.0 (black),
from bottom to top. Plasma-$\beta = 1$ estimates when convective energy transfer is suppressed by 
the magnetic field. Photospheric values are derived from 1D model atmospheres
with a mixing-length parameterization of ML2/$\alpha$ = 0.8 \citep{tremblay11} while values 
for the base of the convection zone are derived from standard envelope models using 
a slightly more efficient ML2/$\alpha$ = 1.0 convection \citep{fontaine01}.
 \label{fg:f_supres}}
\end{figure}

We have employed a standard grid of 1D model atmospheres \citep{tremblay11} to
compute the critical magnetic field strength, defined by $\beta = 1$, above which convection 
is significantly suppressed in the photosphere ($\tau_{R} = 1$). Figure~\ref{fg:f_supres} shows 
that the critical field is always below $\sim$50~kG. Known magnetic white dwarfs have field 
strengths typically much larger than these values, and our results 
suggest that convection is suppressed at the surface
of HFMWDs. Furthermore, while we have only performed simulations with a vertically
oriented magnetic field, it is generally thought that the damping of
convection is even stronger for horizontally oriented fields
since the Lorentz force will act against vertical flows. In other words, 
convection is expected to be globally inhibited above a certain magnetic 
field strength \citep{nature}.

The rapid increase of $\beta$ as a function of depth implies that 
when convection is suppressed at the surface, it could still be 
fully developed in deeper layers as demonstrated by our 5~kG simulation.
Once $\beta$ = 1 at the base of the convection zone, the entire convection zone is likely to 
be significantly disrupted. Figure~\ref{fg:f_supres} shows this critical 
field strength (dashed lines) as predicted by 1D envelopes \citep{fontaine01}.
In the intermediate regime between the suppression
of convection at the surface and in the full convection zone, one 
should use radiative atmospheres but consider the possibility of
an internal convection zone. However, new cooling sequences with partial convective inhibition 
would need to be computed to determine the size and structure of 
these internal convection zones. These calculations are outside the scope of this work
because a realistic magnetic field geometry would be required to
properly model individual white dwarfs. Furthermore, it is difficult to extrapolate our RMHD results 
for the atmosphere, where convective velocities are close
to the sound speed, to deeper convective layers where the
flows have a kinetic energy density that becomes far 
smaller than the thermal energy density. Once the 
magnetic field becomes larger than the kinetic equipartition field strength

\begin{equation}
B_{\rm eq}^2/8\pi = \frac{1}{2} \langle \rho v_{\rm conv}^2 \rangle~,
\end{equation}

\begin{figure}[t]
\begin{center}
\includegraphics[width=0.50\linewidth,bb=110 150 450 680]{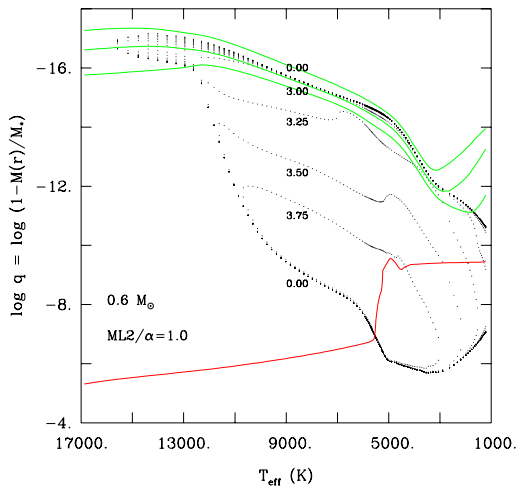}
\end{center}
\caption{Contours of kinetic equipartition magnetic field strength (logarithmic values in Gauss 
identified on the panel, see Eq.~(2)) as a function of fractional mass $\log q = \log(1-M(r)/M_{*})$ integrated from the surface
and $T_{\rm eff}$. We rely on the evolution sequence at 0.6~$M_{\odot}$
with ML2/$\alpha$ = 1.0 convection (see Section 2.2). We also show the position of three
atmospheric layers ($\tau_{\rm R}$ = 0.1, 1.0, 10.0, from top to bottom in solid green lines) and the degeneracy
boundary ($\eta$ = 0, solid red line). 
 \label{fg:f_equip}}
\end{figure}

{\noindent}where $v_{\rm conv}$ is the local convective velocity,
different modes of convection with smaller physical scales may set in.
Figure~\ref{fg:f_equip} demonstrates that the kinetic equipartition field 
strength is in the kG-range throughout the convection zone for a 
representative 0.6~$M_{\odot}$ white dwarf. It suggests that convection
could be disrupted for magnetic field strengths smaller than those defined by the conservative
$\beta = 1$ estimate of Figure~\ref{fg:f_supres} for the bottom of the convection zone.

\subsection{Evolutionary Models}
  
It has been known for a long time that superficial convection has no
influence whatsoever on the cooling time until the base of the
convection zone reaches into the degenerate reservoir of thermal
energy and couples, for the first time in the cooling process, the
surface with that reservoir \citep{tassoul90,fontaine01}. The convective 
coupling occurs at $T_{\rm eff}$ values lower than 6000~K in white dwarfs, 
hence the suppression of convection is not expected to impact cooling
rates for warmer remnants. This argument contradicts the suggestion
from \citet[][see Fig. 3a]{nature} that the suppression of convection changes
the cooling rates and explains the observed temperature distribution of 
magnetic white dwarfs, for which their coolest bin is at $T_{\rm eff} = 6000$~K.
To demonstrate it quantitatively, this section presents new
evolution sequences that we have computed with our state-of-the-art white dwarf
evolutionary code \citep{fontaine01,fontaine13}.
To fully appreciate the results, we also review the important properties of white dwarf cooling in Section 2.3.

\begin{figure*}[t]
\begin{center}
\includegraphics[width=0.43\linewidth,bb=0 140 600 690]{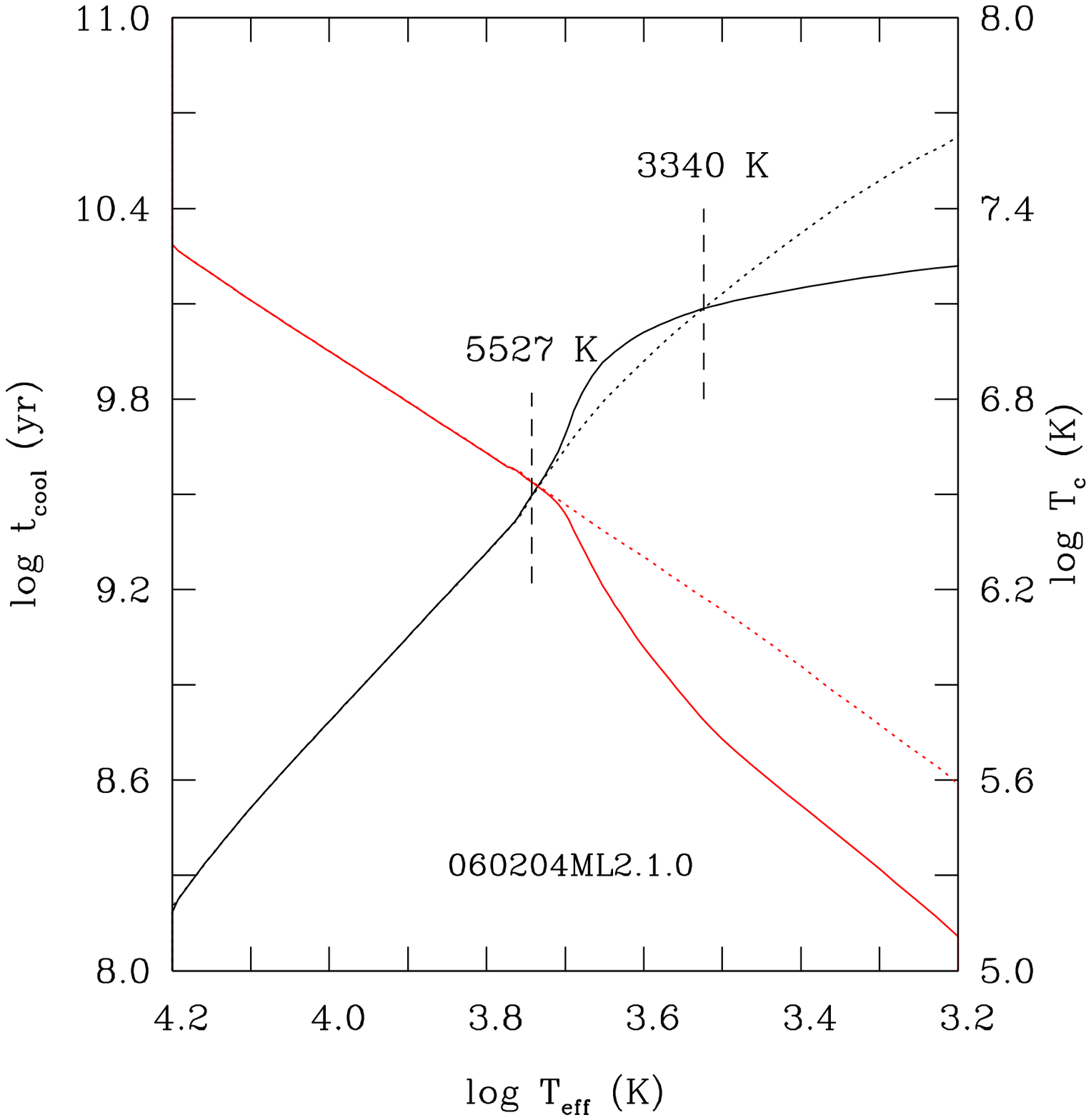}
\includegraphics[width=0.43\linewidth,bb=0 140 600 590]{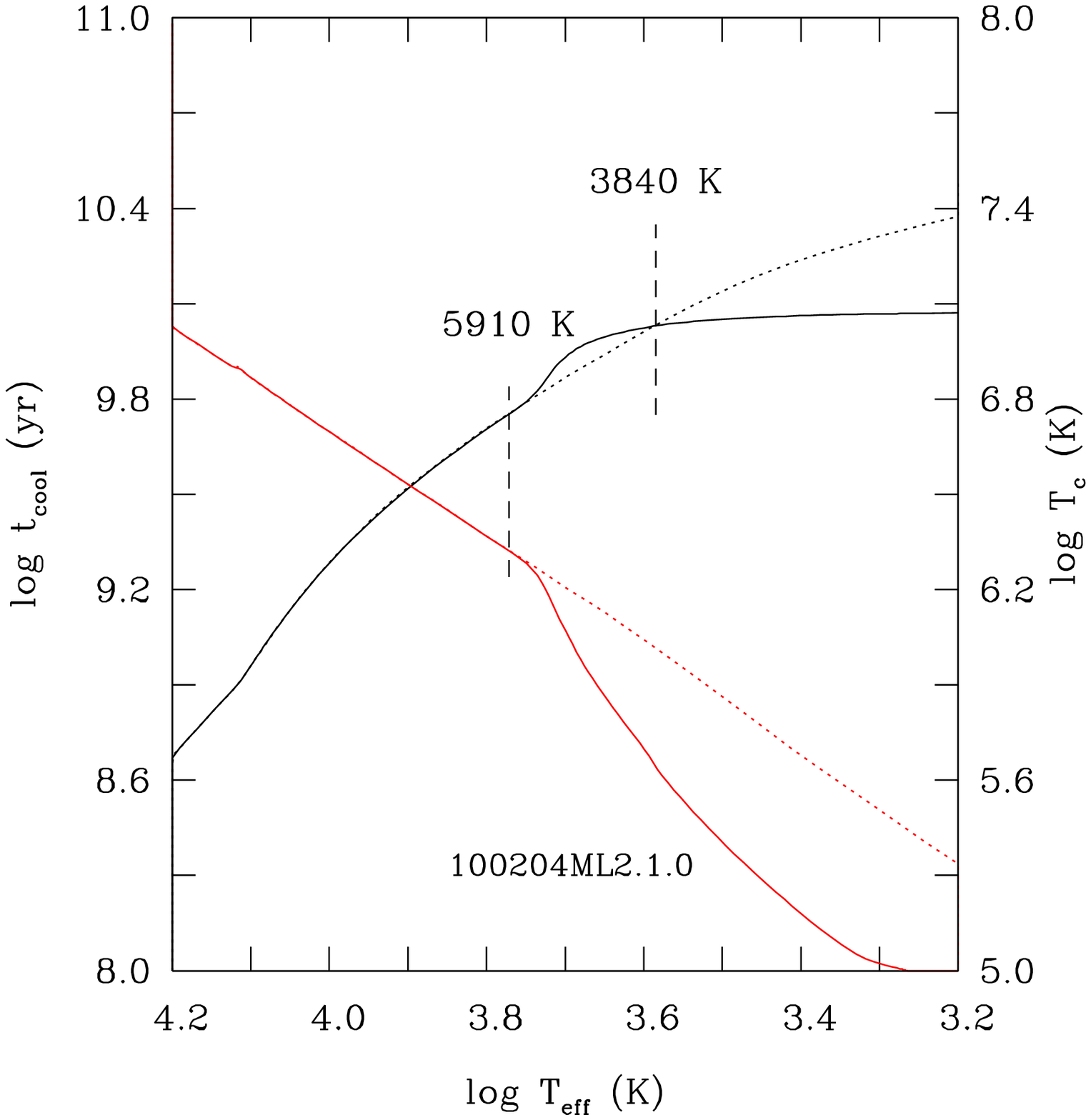}
\end{center}
\caption{{\it (Left:)} cooling sequences in terms of cooling time
  (black, left axis) and central temperature (red, right axis) as a function of decreasing
  $T_{\rm eff}$ for a 0.6 $M_{\odot}$ DA white dwarf (solid
    lines). We have assumed thick helium and hydrogen
  layers with fractional masses of 10$^{-2}$ and 10$^{-4}$,
  respectively. We have also computed a sequence where convection was
  artificially suppressed mimicking the effect of a strong magnetic
  field (dotted lines). Convection has no effect on the cooling until
  there is a convective coupling with the degenerate core at the position 
	illustrated by the first dashed vertical segment from the left. The location
	where the cooling times become larger for the magnetic sequence
  is indicated by the second dashed vertical segment from the left. The
	$T_{\rm eff}$ values for both transitions are shown on the panel.
  {\it (Right:)} same as left panel but for a 1.0~$M_{\odot}$ DA
  white dwarf.
 \label{fg:f_cooling}}
\end{figure*}

We computed a standard 0.6~$M_{\odot}$ sequence with a C/O
core, a helium envelope containing 10$^{-2}$ of the total mass, and a
hydrogen outer layer containing 10$^{-4}$ of the total mass. In particular, it takes into account superficial
convection as it develops with time relying on the so-called
ML2/$\alpha$ = 1.0 version of the mixing-length theory \citep{bohm71,tassoul90}. We have
computed an additional sequence where convection is totally
suppressed, thus mimicking the maximum possible effect of magnetic
inhibition, e.g. for field strengths of 10 MG or larger according
to Figure~\ref{fg:f_supres}. Both sequences are presented in Figure~\ref{fg:f_cooling}
(left panel), where the solid curves refer to the normal sequence,
while the dotted curves refer to the ``magnetic'' sequence. The
location of convective coupling is indicated by the first dashed vertical 
segment from the left. This corresponds specifically to the model with the base of its convection zone first
entering the degenerate thermal reservoir from above (the upper
boundary of that reservoir is defined by a local value of the electron
degeneracy parameter of $\eta$ = 0, where $\eta kT$ is the chemical potential of the free
electrons). When convective coupling occurs,
$T_{\rm eff} =$ 5527~K and the
cooling age is 3.13~Gyr. Above $T_{\rm eff}$ = 5527 K, there is no
significant difference whatsoever between the behaviors of the two
sequences, meaning that magnetic inhibition of superficial convection
does nothing to the cooling process in this hotter phase. We have also computed 
sequences at 1.0 $M_{\odot}$ which are likely more representative of the HFMWDs. 
Figure~\ref{fg:f_cooling} (right) demonstrates that the behavior is similar 
to the lower mass case, and convective coupling takes place at a only slightly 
higher temperature.

\begin{figure}[t]
\begin{center}
\includegraphics[width=0.85\linewidth,bb=0 150 600 700]{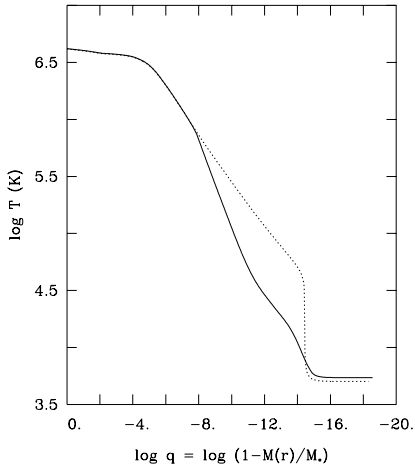}
\end{center}
\caption{Temperature structure as a function of fractional
  mass (both logarithmic values) for a DA white dwarf at $T_{\rm eff} \sim$ 6200~K and 0.6~M$_{\odot}$. The solid sequence ($T_{\rm eff}$ = 6243~K) relies on 1D convection (ML2/$\alpha$ = 1.0)
		while the dotted sequence ($T_{\rm eff}$ = 6205~K) had convection suppressed in the entire cooling process.
 \label{fg:f_t6000}}
\end{figure}

\begin{figure}[t]
\begin{center}
\includegraphics[width=0.85\linewidth,bb=0 150 600 700]{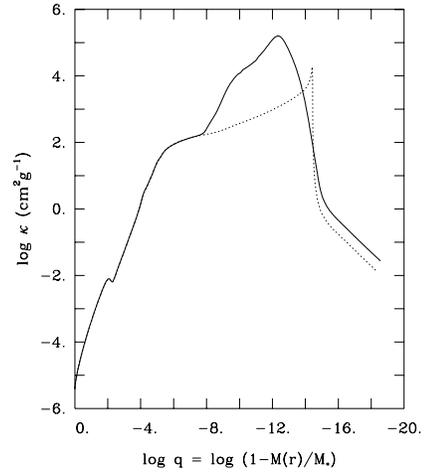}
\end{center}
\caption{Similar to Figure~\ref{fg:f_t6000} but for the Rosseland mean opacity ($\kappa$) as a function of fractional
  mass (both logarithmic values). The solid sequence ($T_{\rm eff}$ = 6243~K) relies on 1D convection (ML2/$\alpha$ = 1.0)
		while the dotted sequence ($T_{\rm eff}$ = 6205~K) had convection suppressed in the entire cooling process.
 \label{fg:f_o6000}}
\end{figure}

Our evolutionary sequences demonstrate that the cooling rates, 
hence the relation between core and surface temperature, 
must remain the same for magnetic and non-magnetic white dwarfs. We now try to reconcile this fact
with the prediction from our RMHD simulations indicating that the inhibition of convection 
by a magnetic field creates a steeper (radiative) temperature gradient in 
the outer convectively unstable layers. Figure~\ref{fg:f_t6000}
presents the temperature profile of a model at $T_{\rm eff} \sim 6200$~K from the standard evolution 
sequence at 0.6~$M_{\odot}$, along with the case where convection was suppressed for the entire cooling
process. It confirms that even though there is a much steeper gradient
at the surface of magnetic white dwarfs, this is not the case for all internal layers,
and the non-magnetic relation between core and (average) surface temperature holds.
Interestingly, Figure~\ref{fg:f_o6000} demonstrates that for the magnetic case, the steep
radiative gradient in the outer layers is associated with a very sharp opacity peak as 
a function of fractional mass. It is unclear
if such opacity peak could generate pulsations in magnetic white
dwarfs, which we discuss in Section 3.4.

\subsection{The Cooling Process in White Dwarfs}

\begin{figure}[t]
\begin{center}
\includegraphics[width=0.85\linewidth,bb=50 150 550 680]{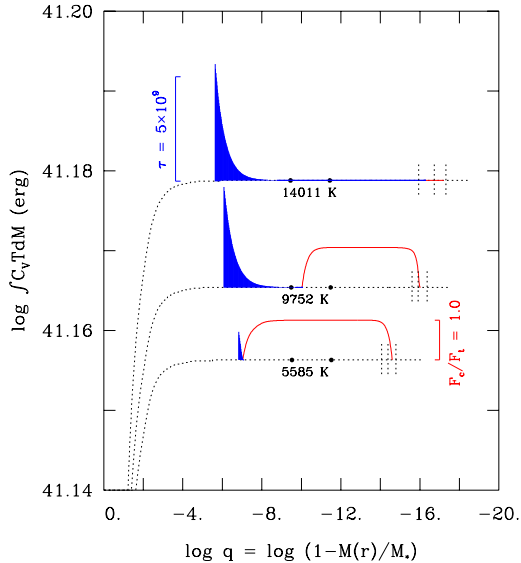}
\end{center}
\caption{Logarithmic value of the available thermal energy integrated from the center (left to right, black dotted curves) at three given epochs of the standard 0.6 $M_{\odot}$ cooling sequence ($T_{\rm eff}$ = 14011~K, 9752~K, and 5585~K, identified on the panel) within a certain fractional mass $\log q$. The uppermost degenerate layer ($\eta = 0$) corresponds, for each model, to the
location of the sharp cutoff on the left of the blue spike. The blue spikes correspond to 
the running integral of the optical depth, from the base of the convection zone on the right to the layer where $\eta = 0$ on the left, with a scale of $\tau_{\rm R} = 5\times10^{9}$.
The ratio of the convective to total flux is illustrated by the red profiles (ML2/$\alpha$ = 1.0) and the two black dots 
on each curve indicate, respectively, the depth where the magnetic pressure is equal to 
the gas pressure assuming a magnetic field of 10~MG (on the left) and 1~MG (on the right). We also indicate
the location of three atmospheric layers along the x-axis, corresponding to $\tau_{\rm R}$
= 10, 1, and 0.1, from left to right (the short vertical dotted line segments). 
 \label{fg:f_cooling2}}
\end{figure}

We have designed Figure~\ref{fg:f_cooling2} to review the cooling process in white dwarfs.
The cooling time depends on the amount of thermal energy contained in the star and the rate with
which this energy is transferred from the thermal reservoir to the
surface. The available thermal energy at a given epoch is given by the
integral shown on the y-axis of Figure~\ref{fg:f_cooling2}. Here, we show the running 
integral (black dotted curve), from the center to the surface, for three
models belonging to the standard (convective) evolutionary sequence at 0.6 $M_{\odot}$ discussed above. 
The x-axis shows the fractional mass
$\log q = \log(1-M(r)/M_{*})$ integrated from the surface. The upper boundary of the reservoir of thermal energy
is a concept that is a bit fuzzy, but it must correspond to a location
on the flat part of each curve, i.e., to a layer above which there is
practically no more contribution to the reservoir. Conveniently, this
boundary is usually defined as the layer where the
degeneracy parameter $\eta = 0$. In Figure~\ref{fg:f_cooling2}, 
the layer $\eta = 0$ corresponds, for each model, to the
location of the sharp cutoff on the left of the blue spike. With
cooling, the boundary $\eta = 0$ moves up toward the surface because the
star becomes globally increasingly more degenerate. 
Moreover, we have illustrated in red the profile of the ratio of
the convective flux to total flux, $F_{\rm conv}/F_{\rm tot}$. 
It should be understood that convective coupling arises when
the base of the convection zone reaches the boundary $\eta = 0$, which is imminent but
has not yet occurred in the coolest model ($T_{\rm eff}$ = 5585~K) shown in the
plot. In this particular evolutionary sequence, convective coupling
occurs when the star has cooled down to the somewhat lower value of $T_{\rm eff}
= 5527$~K.

In a cooling white dwarf, the degenerate core and reservoir
of thermal energy is relatively well insulated by a nondegenerate 
envelope whose global opacity regulates the rate of energy loss. To
illustrate this opacity barrier, and in particular the role of the
insulating layers between the base of the outer convection zone and the
reservoir, we integrated the optical depth ${\rm d}\tau_{\rm R} = -\kappa\rho{\rm d}r$ between the base of the convection zone
and the layer $\eta$ = 0. For each model considered, we plotted in Figure~\ref{fg:f_cooling2} in blue 
the running integral of optical depth, from right to left together 
with a scale of $\tau_{\rm R} = 5\times10^{9}$. The blue spikes thus 
identify the layers that are of importance in the
insulating process and in the role of regulator of the rate of energy
transfer from the core to the surface. Even for the coolest model
illustrated here, the opacity barrier is still enormous and the reservoir
remains relatively well insulated. The convective coupling will occur in a 
somewhat cooler phase for which the
base of the convection zone finally reaches the boundary $\eta = 0$. 
From that point on in time, the reservoir becomes effectively coupled directly 
to the atmospheric layers via a convection zone whose efficiency reaches practically
100\%. For the first time in the evolution of the star, the exact physical conditions 
characterizing the atmospheric layers will start playing a role in the cooling process.

The layers where the blue optical depth curve is flat in Figure~\ref{fg:f_cooling2} have a negligible 
contribution to the opacity barrier and, thus, cannot play any role in the cooling process. For
example, for the two warmest models, all of the layers above $\log
(1-M(r)/M_{*}) \sim -9$ have no impact on this process. For the coolest model,
all of the layers above $\log (1-M(r)/M_{*}) \sim -7$ have no impact either; the
insulating layer represented by the small blue spike is still
extremely efficient at regulating by itself the outflow of energy. 
In this context, we have added in the figure two black dots on each curve which 
indicate, respectively, the depth where the magnetic pressure is equal to 
the gas pressure assuming a magnetic field of 10 MG (on the left) and 1 MG (on the right). These
layers sit far above the opacity barrier, hence magnetic
effects, namely magnetic pressure, may impact the actual stratification 
of these outer layers, but these layers play no role in
the cooling process. They have negligible 
contribution to the energy reservoir and negligible contribution to 
the opacity barrier.

\begin{figure}[t]
\begin{center}
\includegraphics[width=0.50\linewidth,bb=110 150 450 680]{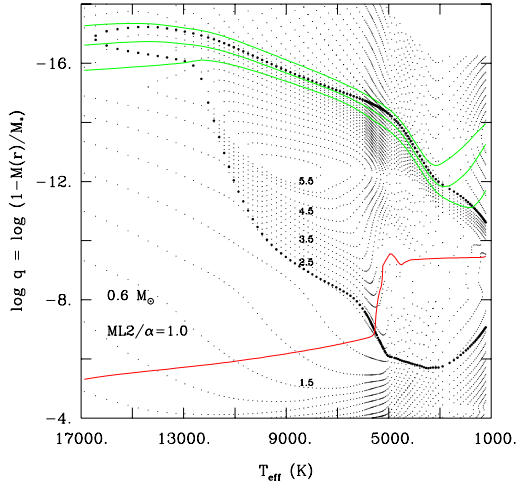}
\end{center}
\caption{Opacity contours (small dots, logarithmic values in cm$^2$ g$^{-1}$ identified on the panel) in the envelope of 
a 0.6~$M_{\odot}$ white dwarf as a function of fractional mass (logarithmic scale) and $T_{\rm eff}$. We rely
on the standard convective sequence with ML2/$\alpha$ = 1.0, and the convection zone is
illustrated by the bold dotted contours. We also show the position of three
atmospheric layers ($\tau_{\rm R}$ = 0.1, 1.0, 10.0, from top to bottom in solid green lines) and the degeneracy
boundary ($\eta$ = 0, solid red line). 
 \label{fg:f_coupling2}}
\end{figure}

A last view on convective coupling can be made from Figure~\ref{fg:f_coupling2}
with the standard convective cooling sequence at 0.6~$M_{\odot}$. The small
dots represent the opacity contours, while the bold dots represent the convective
layers. The opacity maximum is caused by hydrogen recombination. We also show
the position of the degeneracy boundary ($\eta$ = 0) with a solid red curve.
It is observed that when the degeneracy boundary crosses the convection zone,
there is a radical change in the envelope stratification, and conductive
transfer dominates for regions below the degeneracy boundary. Figure~\ref{fg:f_cooling} also
demonstrates that for both the 0.6 and 1.0 $M_{\odot}$ cases, the cooling time of the normal convective sequence 
becomes larger than that of the magnetic sequence in the phase following the onset of convective coupling,
while the central temperature immediately drops below that of the
  magnetic model. This behavior has been explained by
\citet{tassoul90} and \citet{fontaine01}, and it is perhaps best understood with
the analogy of a warm oven. Convective coupling is like opening the
door of the oven; there is initially an excess of heat coming out of
the oven, while the inside temperature drops immediately. In a white
dwarf undergoing convective coupling, the excess of thermal energy is
translated into a delay in the cooling process and the cooling time
increases accordingly. After this excess energy has been radiated
away, convective coupling enters a second phase, and that is that of
accelerated cooling because convection now couples for good the energy
reservoir and the surface, and it transfers energy at a greater rate
than radiation alone could do. It is thus only in this second phase of
the process that the cooling time of the magnetic sequence
becomes larger than the cooling time of the normal sequence, as suggested
by \citet{nature}. In Figure~\ref{fg:f_cooling}, the vertical dashed line segments, marked 
$T_{\rm eff}$ = 3340 and 3840~K for the 0.6 and 1.0~$M_{\odot}$ models, respectively, 
indicate the very low $T_{\rm eff}$ values below which this second phase can proceed. 

We conclude this section with a comparison to the cooling process 
in magnetized neutron stars, which is also 
regulated by a heat blanketing envelope between the atmosphere
and the stellar interior \citep[see, e.g.,][]{neutron1}. For these 
objects, thermal conduction is the dominant energy transfer mechanism
in the degenerate electron gas within the insulating layers, and 
it has been established that the suppression of thermal 
conduction in the direction transverse to the magnetic field lines
can influence the cooling rates \citep{neutron2,neutron3}. 
In a white dwarf, however, the insulating region is non-degenerate
and thermal conduction only takes place in the stellar interior, 
where changes in the conduction rates are unlikely to impact 
the cooling process. Average magnetic fields are also much weaker
in white dwarfs in comparison to magnetized neutron stars.

\subsection{Magnetic Effects on Structures}

\begin{figure}[t]
\begin{center}
\includegraphics[width=0.70\linewidth,bb=50 150 550 680]{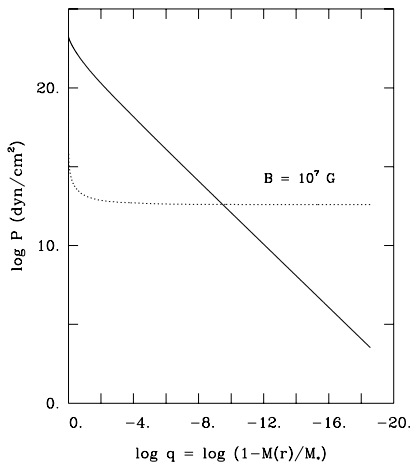}
\end{center} 
\caption{Thermal pressure profile (solid line) of a DA white dwarf structure at 0.6 $M_{\odot}$
and $T_{\rm eff}$ = 9752~K, the same model as on the middle panel of  
Figure~\ref{fg:f_cooling2}. We also show the magnetic pressure (dotted line) for a field strength
of 10~MG at the surface and assuming the conservation of the magnetic flux in the interior. 
 \label{fg:mag_int}}
\end{figure}

Figure~\ref{fg:mag_int} compares the gas and magnetic pressure for 
a characteristic structure at 0.6~$M_{\odot}$ and $T_{\rm eff} \sim 9750$~K. We assume a 
10~MG field at the surface and a conservation of the magnetic flux $4 \pi B r^2$ in the 
interior. This is obviously a rough description of the actual 
magnetic geometry in the interior, which is poorly constrained 
by observations. Nevertheless, it demonstrates that magnetic effects
could only play a role in the outer layers and at the very center, although 
there is no evidence that magnetic field lines reach the central region. 
For the illustrated model, a fractional mass depth of $\log (1-M(r)/M_{*}) = -9$ 
corresponds approximately to a fractional radius of $\log (1-r/R_{*}) = -2.3$. Thus,
magnetic fields (at the 10 MG level) could at best only in the outermost 0.5\%
of the radius have an influence on the structure of these representative white dwarf
models. As a consequence, we conclude
that current mass-radius relations for non-magnetic white dwarfs will hold
for magnetic remnants as well.

\section{DISCUSSION}

The computation of RMHD simulations for DA white dwarfs confirms that
convective energy transport is seriously impeded by magnetic field lines when
the plasma-$\beta$ parameter is smaller than unity. As a
consequence, radiative 1D model atmospheres can be employed for magnetic white 
dwarfs with $B \gtrsim$ 50 kG 
according to Figure~\ref{fg:f_supres}. The main shortcoming
in the modeling of most known magnetic white dwarfs remains the 
spectral synthesis of the Balmer lines accounting for both Stark and 
Zeeman effects \citep{wickra86}.

\subsection{Photometric Variability of Magnetic White Dwarfs}

It is difficult to explain from our results the large number of magnetic white dwarfs 
that show photometric variations of a few percent 
over their rotation period \citep{brinkworth13,lawrie13,nature}.
We have demonstrated in Section 2.2 that the partial or total 
suppression of convection is unable to change the 
average surface temperature until there is coupling between the convection zone and the degenerate core
at low $T_{\rm eff}$ values. As a consequence, we can not naturally explain global emergent intensity
variations for the $T_{\rm eff}$ values of known magnetic 
white dwarfs. However, we note that if the magnetic field is moving at the surface,
as hypothesized by \citet{valyavin11} for WD~1953$-$011, the envelope would
take some time to adjust to the new surface conditions. The Kelvin-Helmholtz timescale
of the portion of the envelope including the entire convection zone is one estimate for this thermal relaxation time, which varies from about one second at 12,000~K to about 1000~yr at 6000~K. Since the cooling rates
must remain constant according to our evolutionary models, the flux fluctuations created from
this mechanism would average out over the full surface but not necessarily over the apparent
stellar disk.

We note that photometric variations are observed in hot magnetic white dwarfs 
where no convection is predicted, hence it is already clear that 
convective effects are not involved in some cases.
Previously supplied explanations for photometric variations remain valid, 
such as magneto-optical effects involving radiative transfer under different
polarizations \citep{martin79,wickra86,ferrario97}.
Finally, variations are also observed, although with a weaker
amplitude, in apparently non-magnetic white dwarfs, where accretion hot spots 
or UV flux fluctuations and fluorescent optical re-emission have been suggested
as possible explanations \citep{maoz15}.

\subsection{Cooling Age Distribution of Magnetic White Dwarfs}

Our results do not support the hypothesis that the observed distribution of HFMWDs as a
function of $T_{\rm eff}$ can be explained by different cooling timescales between magnetic and non-magnetic white dwarfs. This does not imply that the number ratio of magnetic to non-magnetic remnants should be constant
as a function of $T_{\rm eff}$. The cooling age distribution of HFMWDs could be
different from the fact alone that they have a distinct mass distribution. A variation of the velocity distribution as a function of both mass and $T_{\rm eff}$ \citep{merger2}, a consequence of the different main-sequence lifetimes, could change the magnetic incidence as a function of $T_{\rm eff}$ even for volume-complete samples. Furthermore,
a distinction between magnetic and non-magnetic objects could be present if a significant fraction of magnetic white dwarfs
originate from mergers, which presumably have a different cooling history compared to single remnants. Finally, very cool DA white dwarfs 
have deep convection zones, and for $T_{\rm eff}
\lesssim 6000$~K, they reach a regime where the convective turnover
timescale at the base of the convection zone is of the order of
a few hours, which is similar to the rotation periods of magnetic white 
dwarfs \citep{brinkworth13}. The hypothesis of a $\alpha\omega$ convective dynamo
becomes tantalizing, although this needs to be tested with dynamical models. However, this
dynamo is unlikely to generate fields stronger than the kinetic equipartition field strength \citep{fontaine73,thomas95,dufour08}. Figure~\ref{fg:f_equip} 
demonstrates that for our standard evolutionary sequence at 0.6~$M_{\odot}$,
the equipartition field strength reaches a maximum value
of $B_{\rm eq} \sim 10$~kG at the base of the convection zone, suggesting
it is an unlikely scenario for the known magnetic white dwarfs.

We have found no firm evidence in the literature for a variation in the incidence of magnetic
white dwarfs as a function of $T_{\rm eff}$, which differs from the claim of \citet{nature} 
that the picture has now been settled. On the contrary, \citet{liebert03}, \citet{hollands15}, and \citet{ferrario15} suggest that variations still
need to be confirmed owing to several observational biases and conflicting results. Furthermore, 
\citet{kulebi09} and \citet{kepler13} 
find no clear evidence of variations in the homogeneous SDSS sample, although most objects have $T_{\rm eff} > 7000$~K, above the temperature where
\citet{nature} observe a significant increase. There is marginal evidence from the local 20 pc sample \citep{20pc} 
that the incidence of magnetic fields increases for $T_{\rm eff} <$ 6000~K.
If we consider only DA white dwarfs as well as objects
with a derived distance under 20 pc in Table~2 of \citet{20pc},
we find a magnetic incidence of $22 \pm 11$\% (4 magnetic objects) for 5000 $< T_{\rm eff}$ (K) $<$ 6000,
while the value is $10 \pm 5$\% for warmer objects.
We believe it is necessary to confirm this behavior with larger samples
to fully understand the evolution of magnetic white dwarfs.

\subsection{Magnetic Fields in the White Dwarf Population}

Few magnetic white dwarfs have precise atmospheric parameter
determinations, and it is typical to exclude them from the samples employed to derive the mean
properties of field white dwarfs \citep[see, e.g.,][]{tremblay11}. It is however difficult to detect 
magnetic objects with $B \lesssim$ 1 MG
at low spectral resolution, hence it is therefore nearly impossible to
define clean non-magnetic samples. 

We have shown that magnetic fields of a few kG can significantly
impact the thermal stratification in the upper layers of convective
DA white dwarfs. Yet these fields are too weak to produce any significant 
Zeeman splitting, hence white dwarfs harboring such fields would 
not easily be detected. \citet{kepler13} have suggested that
undetected magnetic fields could explain the so-called high-$\log g$ problem
observed in the white dwarf mass distributions \citep{bergeron90}.
On the other hand, it was recently demonstrated that this problem is
instead caused by inaccuracies in the 1D mixing-length convection
model \citep{tremblay13}. 
Furthermore, \citet{kepler13} suggest that field strengths increase for convective objects,
which would be a manifestation of the amplification of magnetic fields by
convection. However, all their observations have $B > 1$~MG, 
which is too strong to be amplified by convection since the kinetic 
equipartition field strength is always much smaller than $B = 1$~MG 
as demonstrated in Figure~\ref{fg:f_equip}.

\begin{figure}[t]
\begin{center}
\includegraphics[width=0.47\linewidth,bb=100 135 490 735]{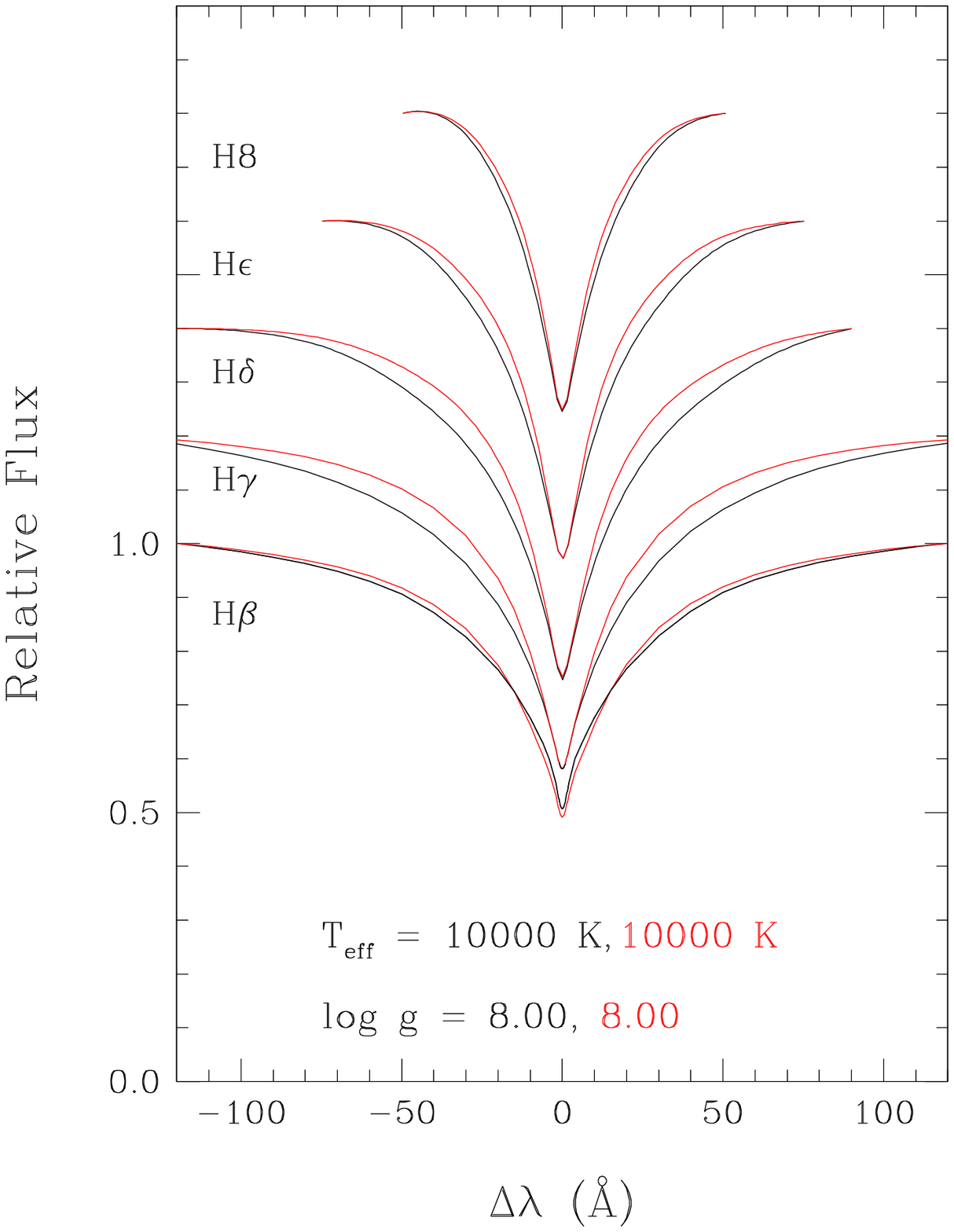}
\includegraphics[width=0.47\linewidth,bb=100 135 490 735]{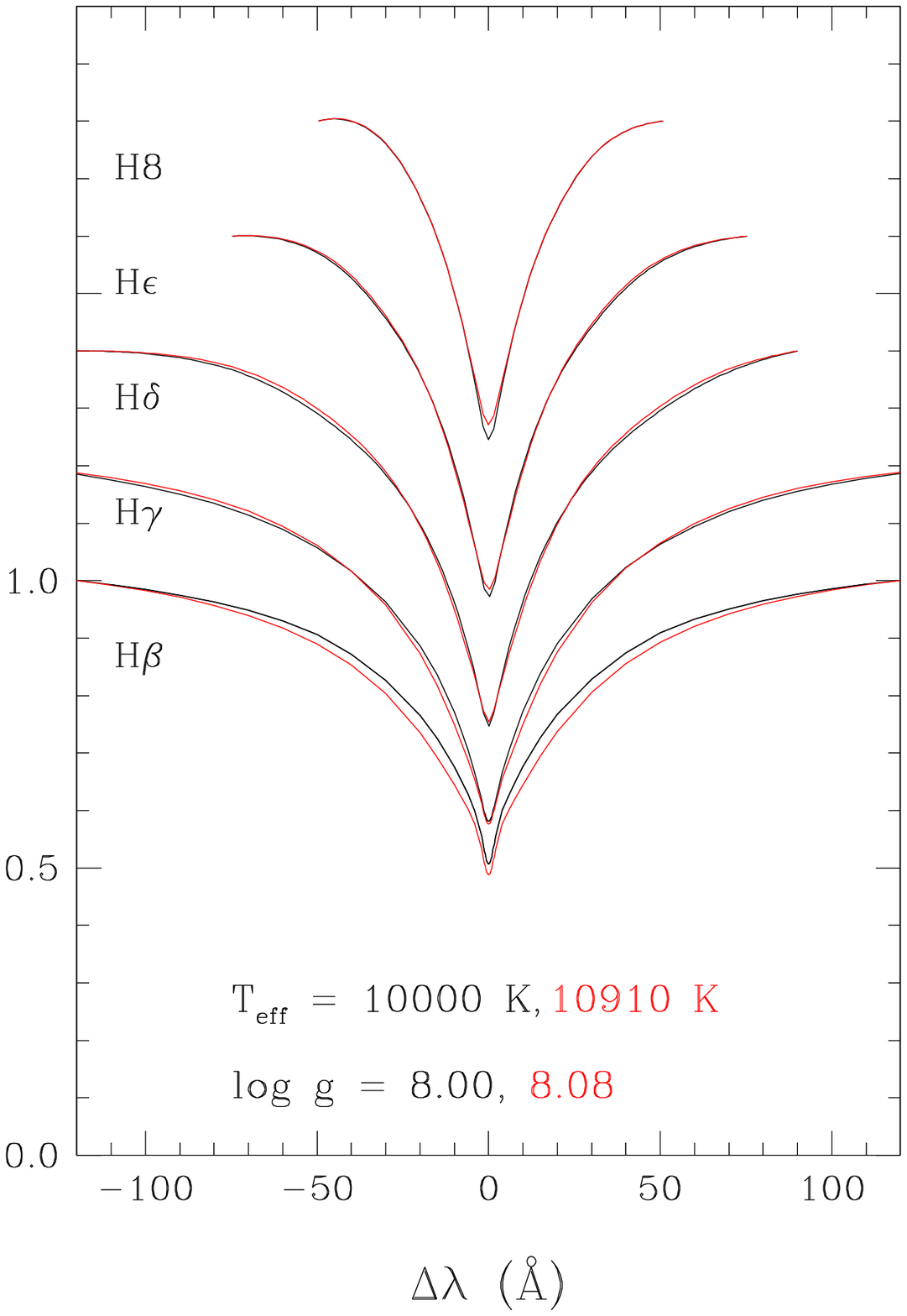}
\caption{{\it Left:} Predicted Balmer line profiles for a DA white dwarf at $T_{\rm eff} = 10,000$~K
and $\log g =$ 8.0. The spectra were computed from a standard 1D model atmosphere with ML2$/\alpha$ = 0.8
convection (red) and a radiative 1D atmosphere where convection was artificially 
suppressed (black), representing the effect of a $B \gtrsim 5$~kG magnetic field. 
All profiles are normalized to a unity continuum and the transitions are identified
on the panel. We have employed a convolution of 3~\AA~(FWHM) to represent typical observations. {\it Right:}
Similar to left panel but we show the standard convective 1D model spectra 
(red, $T_{\rm eff} = 10,910$~K and $\log g = 8.08$) that best fits the radiative model. 
It suggests that fitting this magnetic white dwarf with a proper radiative model would result 
in $\Delta \log g = -0.08$ and $\Delta T_{\rm eff} = -$910~K compared to the standard 
convective solution. \label{fg:f_spec_noconv}}
\end{center}
\end{figure}

To understand the effects of a population of white dwarfs
with small undetected magnetic fields, we have computed 
synthetic 1D spectra at $T_{\rm eff} = 10,000$~K 
and $\log g$ = 8.0. The spectra are derived from both a standard convective model atmosphere with 
ML2/$\alpha = 0.8$ \citep{tremblay11}, and a radiative atmosphere where convection 
was completely inhibited, mimicking the effect of a weak $5 \lesssim B$ (kG) $\lesssim 100$ magnetic field, 
i.e. the range where Zeeman splitting is negligible at low spectral resolution. The left panel of 
Figure~\ref{fg:f_spec_noconv} demonstrates that the predicted Balmer
lines of the two models are significantly different, although when projecting the magnetic model on a grid of convective models on the right panel of Figure~\ref{fg:f_spec_noconv}, the Balmer
lines look much alike, albeit with an offset in the atmospheric parameters. It implies that it would be difficult to identify such a small magnetic field from spectroscopy alone. This could 
have an impact on the observed mass distribution of cool convective white dwarfs,
although the $\log g$ shift is moderate 
according to Figure~\ref{fg:f_spec_noconv}, and the incidence of magnetic white 
dwarfs in the $\sim$ 10 kG-range is expected to be small \citep{kawka12,landstreet12}.

The situation is different when accounting for 3D effects. In that more realistic case, the magnetic field inhibits convective overshoot so that the upper layers ($\tau_{\rm R} \lesssim 10^{-2}$) must be in radiative
equilibrium (see Figure~\ref{fg:f_struc}). In 1D, these upper layers are never convective
and always in radiative equilibrium. As a consequence, synthetic spectral line cores based on 3D simulations are significantly
shallower in the magnetic case, while they do not change in 1D. 
We refrain from a quantitative prediction at this stage since the 3D RMHD structures have been
computed with different numerical parameters in comparison to
the published 3D grid. Nevertheless, it is a potential 
explanation for the problem observed in \citet{tremblay13},
where the predicted 3D line cores are systematically
too deep and suggest that the upper layers are too cool. 
Figure~\ref{fg:f_struc} illustrates that a field strength of $\sim$ 1~kG is sufficient
to significantly heat the upper layers.

It is unlikely that the commonly proposed evolution scenarios for magnetic white dwarfs could 
systematically generate $\sim$ 1~kG magnetic fields, which would then impact
the observed line cores. A plausible alternative, however, is that a turbulent dynamo systematically 
generates weak magnetic fields in convective white dwarfs, a well discussed scenario for quiet regions
of the Sun \citep{cattaneo99,vogler07,moll11}. It consists of the
amplification of small seed magnetic fields by the electrically
conducting turbulent convective flows at the surface. Such fields are likely
to reach an equilibrium strength of a fraction of the 
kinetic equipartition energy, corresponding to 0.1-1 kG in the photosphere
of convective DA white dwarfs according to Figure~\ref{fg:f_equip}. The magnetic fields would have characteristic
dimensions of the convective eddies of at
most a few hundred meters, hence it would be difficult to detect them, except from their
systematic feedback effect on the atmospheric stratification. As a consequence, recent
spectropolarimetric surveys provide no direct constraint on this scenario.
We hope to compute turbulent dynamo
RMHD simulations in the future to verify whether the magnetic fields
reach a sufficient amplitude to solve the discrepancy between
the predicted 3D line cores and observations.

\subsection{Pulsating White Dwarfs}

It is difficult to apply our results quantitatively to pulsating 
white dwarfs. The base of the convection zone corresponds to
the driving region of the ZZ Ceti pulsations \citep[see, e.g.,][]{fontaine08},
hence the dashed lines in Figure~\ref{fg:f_supres} illustrate the critical field where 
convective energy transfer will be largely suppressed in these layers. Thus, magnetic fields stronger than 1~MG will likely have a
dramatic effect on the driving mechanism of the pulsations, although it is difficult 
to rule out pulsating instabilities at this stage since the stratification 
will still be unstable. Another aspect of the problem is that the
inhibition of convection will create a strong temperature gradient and
opacity peak in the convectively unstable upper layers (see Figure~\ref{fg:f_o6000}), which could independently drive
pulsations through a $\kappa$-type mechanism. This process has already
been suggested for pulsating and strongly magnetic hot DQ white dwarfs
\citep{dufour08}. 

It is difficult to predict the position of an
instability strip for magnetic white dwarfs since it is likely to depend on the strength and 
geometry of the magnetic fields. Indeed, magnetic pressure will impact 
the position of the opacity peak as function of the radius. We note
that the Lorentz force affects nonradial pulsations as well \citep{saio13},
requiring additional theoretical work to model pulsating magnetic white dwarfs.
However, the Ohmic timescale in the outer layers is short, suggesting that the 
magnetic field could be relaxed to a force-free potential state. This further
highlights the fact that one must rely on realistic magnetic field 
geometries to model pulsations in magnetic white dwarfs.

For DA atmospheres, MG-range fields are
excluded for the 56 bright ZZ Ceti white dwarfs in the
\citet{gianninas11} sample, suggesting that magnetic fields inhibit
pulsations. On the other hand, none of the HFMWDs with known $T_{\rm eff}$
and $\log g$ \citep{briggs15} are within the ZZ Ceti instability
strip, an essential ingredient to conclude about the possibility of
HFMWD ZZ Ceti white dwarfs. 

\section{CONCLUSION}

We have computed the first RMHD simulations of pure-hydrogen white
dwarf atmospheres. We have demonstrated that convective energy
transfer is largely suppressed in the atmosphere of magnetic 
white dwarfs for field strengths larger than $B \sim 50$~kG, confirming
quantitatively the previously widespread idea that HFMWDs have
no surface convection. Stronger magnetic fields are necessary to fully
suppress convection in the envelope, and we find that for
$B$ = 1-100~MG, depending on the atmospheric parameters, 
the full stratification becomes radiative. For intermediate field strengths, 
the suppression of convection in the upper layers will change
the stellar structure in a complex way, and new calculations
with partial convective inhibition and realistic magnetic field configurations
must be performed to better understand these objects.

We have presented new evolutionary calculations for DA white
dwarfs where convection was fully suppressed, e.g. mimicking 
the effect of a $B \gtrsim 10$ MG field. We find that the
suppression of convection has no impact on the cooling rates
until there is a convective coupling between the convection zone and the degenerate core 
in the standard sequence at $T_{\rm eff} \sim$ 5500~K. 
The currently known magnetic remnants, which are almost all above this temperature,
are thus cooling like non-magnetic white dwarfs. 
Our results also suggest that the effect of magnetic pressure on the mass-radius 
relation is at most of the order of 1\%. Finally, we conclude that
the photometric variations observed in a large fraction of magnetic white dwarfs
remain largely unexplained. 

\acknowledgements

Support for this work was provided by NASA through Hubble Fellowship
grant \#HF-51329.01 awarded by the Space Telescope Science Institute,
which is operated by the Association of Universities for Research in
Astronomy, Inc., for NASA, under contract NAS 5-26555. This was
supported by the NSERC of Canada and the Canada Research Chair Program.
It was also supported by Sonderforschungsbereich SFB 881 "The Milky Way System"
(Subprojects A4) of the German Research Foundation (DFG).

\end{document}